\documentclass[10pt,twocolumn,letterpaper]{article}

\usepackage{iccv}
\usepackage{times}
\usepackage{epsfig}
\usepackage{graphicx}
\usepackage{amsmath}
\usepackage{amssymb}
\usepackage{subfigure}
\usepackage{xcolor}
\usepackage{multirow}

\usepackage[breaklinks=true,bookmarks=false]{hyperref}

\iccvfinalcopy 

\def\iccvPaperID{6221} 

\ificcvfinal\fi

\begin{document}

\title{Wavelet Domain Style Transfer for an Effective Perception-distortion Tradeoff in Single Image Super-Resolution}

\author{Xin Deng\\
Imperial College London\\
{\tt\small x.deng16@imperial.ac.uk}
\and
Ren Yang\\
ETH Zurich\\
{\tt\small reyang@ee.ethz.ch}
\and
Mai Xu\\
Beihang University\\
{\tt\small maixu@buaa.edu.cn}
\and
Pier Luigi Dragotti\\
Imperial College London\\
{\tt\small p.dragotti@imperial.ac.uk}
}

\maketitle
\ificcvfinal\thispagestyle{empty}\fi

\begin{abstract}
   In single image super-resolution (SISR), given a low-resolution (LR) image, one wishes to find a high-resolution (HR) version of it which is both accurate and photo-realistic.  
   Recently, it has been  shown that there exists a fundamental tradeoff between low distortion and high perceptual quality  \cite{blau2018perception}, and the generative adversarial network (GAN) is demonstrated to approach the perception-distortion (PD) bound effectively. 
   In this paper, we propose a novel method based on  wavelet domain style transfer (WDST), which achieves a better PD tradeoff than the GAN based methods.  Specifically, we propose to use 2D stationary wavelet transform (SWT) to decompose one image into low-frequency  and high-frequency sub-bands. For the low-frequency sub-band, we improve its objective quality through an enhancement network. For the high-frequency sub-band, we propose to use WDST to effectively improve its perceptual quality. By feat of the perfect reconstruction property of wavelets, these sub-bands can be re-combined to obtain an image which has simultaneously high objective and perceptual quality.  The numerical results on various datasets show that our method achieves the best trade-off between the distortion and perceptual quality among the existing state-of-the-art SISR methods.
\end{abstract}
\vspace{-1.2em}
\section{Introduction}
\vspace{-.3em}
Single image super-resolution (SISR) aims to restore a high-resolution (HR) image from a  low-resolution (LR) one. In this context,  some methods focus on improving the objective image quality, through minimizing the mean squared error (MSE) between the restored and the ground-truth images \cite{dong2014learning,shi2016real,kim2016accurate,kim2016deeply,lim2017enhanced,zhang2018image,zhong2018joint}.  Other methods aim to improve the perceptual image quality, through minimizing the perceptual loss using adversarial training \cite{ledig2017photo,sajjadi2017,mechrez2018learning}. The methods driven by objective quality can achieve low distortion but with poor perceptual quality, while the other category can generate photo-realistic images but with large MSE distortion. We wish to obtain a super-resolved  image which is both accurate and photo-realistic. However, as pointed out in \cite{blau2018perception}, there exists a  tradeoff between the ability to achieve low MSE and high perceptual quality.

\begin{figure*}  
	\centering\includegraphics[width=6.5in]{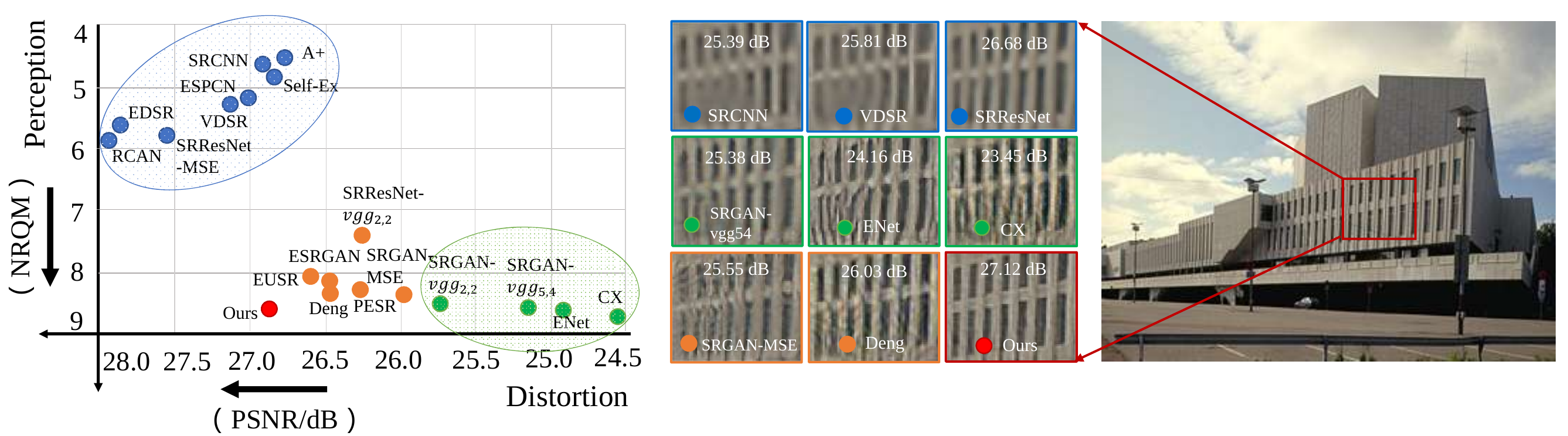}	
	\caption{Perception-distortion performance of different SISR methods. The blue points represent methods aiming for objective quality, the green points represent methods aiming for perceptual quality, and the orange points represent methods aiming for a trade-off between perception and distortion. The higher PSNR value indicates better objective quality and the higher perceptual score  indicates better perceptual quality. The bottom left corner is the best. Our method achieves the best trade-off among all the "orange" methods.}
	\label{comp_build}	
	\vspace{-1em}
\end{figure*}

A natural approach to achieve this tradeoff is to train a generative adversarial network (GAN)  to minimize a combined MSE and adversarial loss, which has been tried by both SRGAN-MSE \cite{ledig2017photo} and ENet \cite{sajjadi2017}. However,  the training process is extremely unstable. On the one hand, the adversarial loss encourages the synthesis of high-frequency details in the results \cite{sajjadi2017}. On the other hand, since these high-frequency details are not in the right place,  the MSE distortion is increased. This unstable training may lead to many undesirable artifacts in the restored image, as shown in Fig. \ref{comp_build}.  To avoid this, ESRGAN \cite{wang2018esrgan}, which is the winner of the PIRM challenge \cite{blau20182018}, proposed to train two separate networks with the low MSE and high perceptual quality targets, respectively.   The two networks are then interpolated to achieve a compromise on the objective and perceptual quality. However, the network interpolation requires that the two networks have exactly the same  architectures, which strongly limits their performance.  Instead of the network interpolation, the image fusion method can be more flexible, since it has no constraint on the network structure. Given one image with high objective quality  and another image with high perceptual quality, image fusion aims to fuse them to obtain an image with both high objective and perceptual quality. 
Recently, Deng \cite{deng2018enhancing} proposed to  combine the two images using image style transfer. However, since the style transfer is performed in pixel domain, it is difficult to preserve the structure and texture information. As shown in Fig. \ref{comp_build}, the structure of the wall is severely affected.  

Another disadvantage of Deng \cite{deng2018enhancing}  is that it  tries to optimize the objective and perceptual quality as a whole, but the objective and perceptual quality are influenced by different elements in an image. When they are optimized as a whole, the increase of objective quality may lead to the decrease of perceptual quality, and vice versa. To achieve the best tradeoff, we should separate the elements affecting the objective quality from those affecting the perceptual quality, and optimize each of them separately. 
In this paper, we propose to use wavelet transform
to achieve this separation, since wavelet can split an image into one low-frequency and several high-frequency sub-bands. We find that the low-frequency sub-band plays an important role in the objective quality, while the high-frequency sub-bands can affect the perceptual quality significantly. After separation, to obtain the best tradeoff,  we use an enhancement network to improve the objective quality of the low-frequency sub-band, and   wavelet domain style transfer to improve the perceptual quality of the high-frequency sub-bands. 

Note that in this paper, we are not aiming for a new SISR
method towards high perceptual or objective image quality,
which has been extensively explored recently. Instead, we propose a
novel image fusion method which combines two images to achieve
the best tradeoff between the perception and distortion, as shown in Fig. \ref{comp_build}.
Our method overcomes many drawbacks of the existing methods.  For example, compared with SRGAN-MSE \cite{ledig2017photo}, we do not need to train a deep network, and thus we have no concerns on the stability of training. Compared with ESRGAN \cite{wang2018esrgan}, we  are more flexible with the choice of the network architecture, which gives us more freedom to achieve the best PD trade-off. Compared with Deng \cite{deng2018enhancing}, we split the elements affecting the objective quality from those affecting the perceptual quality, and we perform the style transfer in the wavelet domain with new techniques. All these contribute to higher reconstruction performance and  a better PD tradeoff.

The main contributions of this work are as follows:
\begin{itemize}
	\item  We show the relationship between  the objective/perceptual image quality and the wavelet sub-bands, which lays an important foundation to push forward the PD performance. 
	Through the wavelet separation, the objective and perceptual quality is allowed to be enhanced separately, with little influence on the other, which leads to a better PD tradeoff. 
	\item We propose a wavelet domain style transfer (WDST) algorithm with a new defined loss function, to achieve an effective tradeoff between distortion and perception.  To the best of our knowledge, we are the first to apply style transfer in the wavelet domain towards a good
	PD tradeoff in SISR. 
	\item We test the performance of our method on various datasets. Compared with other state-of-the-art methods, our method achieves a better tradeoff between the objective and perceptual quality.
	
\end{itemize}

\section{Related work} \label{rl}

\textbf{SISR methods for objective quality.} To improve the objective quality, most methods try to minimize the MSE loss between the reconstructed image and the ground-truth.  Traditional methods rely on dictionary learning  to learn the mapping from LR patches to HR patches \cite{yang2010image,zeyde2010single,timofte2014a+}. The state-of-the-art methods trained a specially-designed deep neural network to minimize the MSE loss between the LR and HR images\cite{dong2014learning, kim2016accurate,shi2016real,lim2017enhanced,huang2017wavelet,zhong2018joint,zhang2018image}. This kind of methods can generate HR images with high objective quality. However, these images are often visually unpleasant with blurred edges, due to the absence of high-frequency details, especially for large upscaling factors.

\textbf{SISR methods for perceptual quality.} Since the MSE loss cannot measure the perceptual similarity between two images, Ledig \textit{et.al} \cite{ledig2017photo} proposed to minimize the perceptual loss which was defined as a weighted sum of VGG loss and adversarial loss.  The VGG loss is good at representing  the perceptual similarity between two images, and the adversarial loss can make the restored image look realistic. Later,  Saggadi  \textit{et.al} \cite{sajjadi2017} proposed to add a texture matching loss to the VGG loss and adversarial loss, which achieved good results in reconstructing images with high perceptual quality.  Recently, Mechrez \textit{et.al}  \cite{mechrez2018learning} proposed the contextual loss to make the internal statistics of the restored image similar to the ground-truth, which leads to more realistic images. 

\textbf{SISR methods for tradeoff between objective and perceptual quality.} Both \cite{ledig2017photo} and \cite{sajjadi2017} have tried to optimize the objective and perceptual quality simultaneously. Specifically, in \cite{ledig2017photo}, the SRGAN-MSE method is proposed to minimize the combined loss of MSE and adversarial losses.  In \cite{sajjadi2017}, another texture matching loss is added to the MSE and adversarial loss to make the training process more stable. However, their results still suffer from  blocking and noisy artifacts. Choi \textit{et.al} \cite{choi2018deep} trains a multi-scale super-resolution model  with a discriminator network and two qualitative score predictors, which achieves high perceptual quality while preserving the objective quality.  Most recently, ESRGAN \cite{wang2018esrgan}  proposed to train two networks which aim to enhance the objective and perceptual quality, respectively, and then these two networks are interpolated to achieve a tradeoff between the objective and perceptual quality.
The work most  related with ours is \cite{deng2018enhancing}, which also uses style transfer to combine two images. However, in \cite{deng2018enhancing}, the style transfer algorithm is performed in the pixel domain, and it has no technique to split the objective and perceptual quality related elements from each other.  As a result, the objective and perceptual quality are optimized as a whole, which significantly decreases the perception-distortion performance. 

\vspace{-.3em}
\section{Proposed method} \label{pm}
\begin{figure}  
	\centering\includegraphics[width=3.3in]{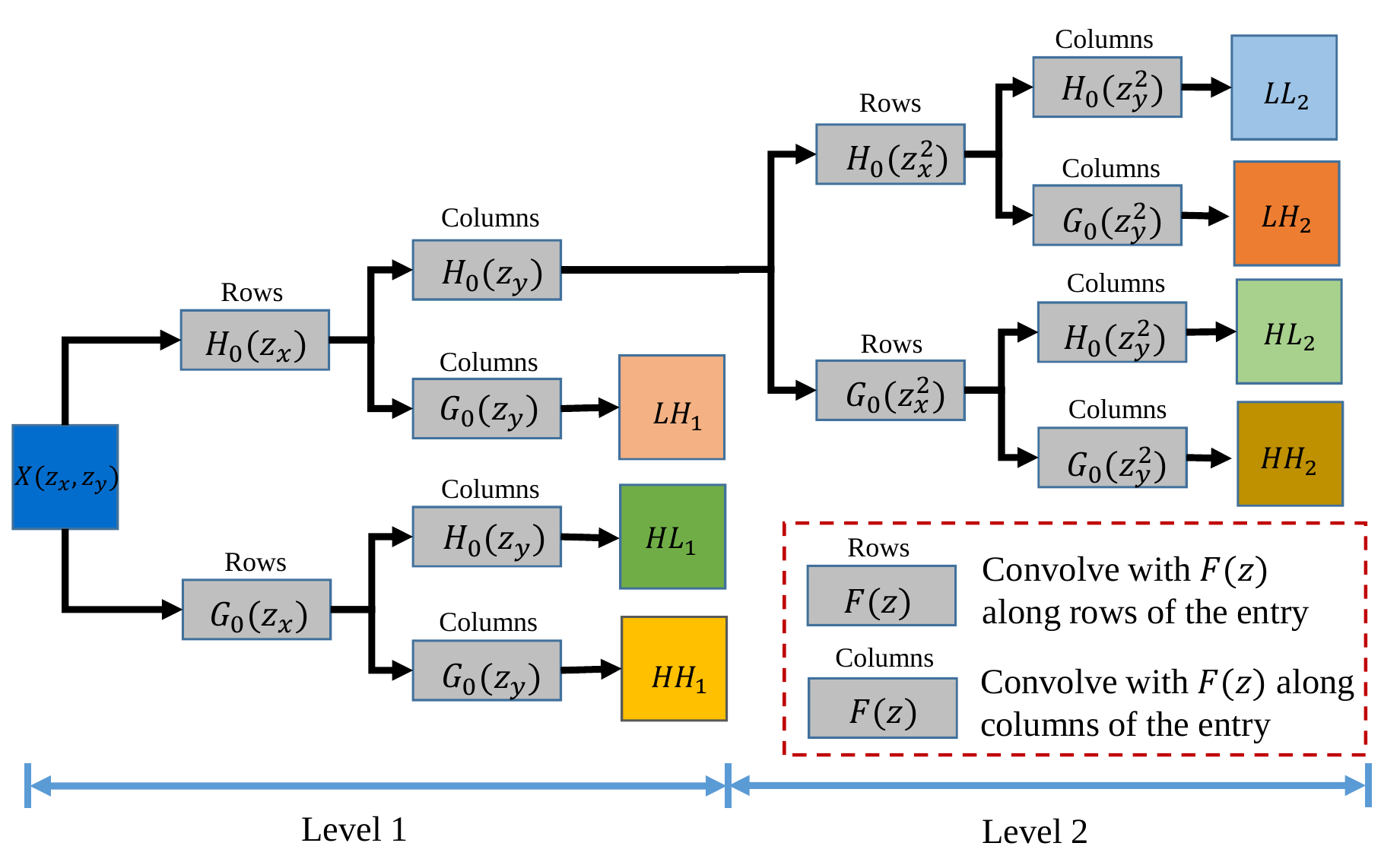}
	\caption{Illustration of two level 2D stationary wavelet transform (SWT) of image $X$, with $H_0$ and $G_0$ as the low-pass and high-pass filters, respectively.  }
	\label{swt}
	\vspace{-1.5em}
\end{figure}

\vspace{-.2em}
\textbf{Stationary wavelet transform. } The wavelet transform allows the multi-resolution analysis of images \cite{jawerth1994overview}. The classical discrete wavelet transform (DWT) has a drawback, i.e., it is not shift-invariant.  The stationary wavelet transform (SWT), also known as undecimated wavelet transform,  overcomes this drawback by removing the downsampling operation in DWT \cite{starck2007undecimated}. Fig. \ref{swt} illustrates the 2D SWT process for 2 level decomposition. Suppose that $H_0$ and $G_0$ are the low-pass and high-pass filters of a standard 1D wavelet decomposition, we can obtain the $z$ transform of $LL$, $LH$, $HL$, and $HH$ sub-bands at the $i$-th level through the following formulations:
\begin{eqnarray} \label{dre}
\small
LL_i(z_x,z_y)=H_0(z_y^{2^{i-1}})H_0(z_x^{2^{i-1}})LL_{i-1}(z_x,z_y), \\
LH_i(z_x,z_y)=G_0(z_y^{2^{i-1}})H_0(z_x^{2^{i-1}})LL_{i-1}(z_x,z_y), \\
HL_i(z_x,z_y)=H_0(z_y^{2^{i-1}})G_0(z_x^{2^{i-1}})LL_{i-1}(z_x,z_y), \\
HH_i(z_x,z_y)=G_0(z_y^{2^{i-1}})G_0(z_x^{2^{i-1}})LL_{i-1}(z_x,z_y), 
\end{eqnarray}
where the $LL_{i-1}$ is the LL sub-band at the $(i-1)$-th level, with $LL_0$ as the input image $X$.  
After the $N$-th level decomposition,  we  obtain (3$N$+1) wavelet sub-bands with the same size as the input image, i.e., $LL_N, \{LH_i\}_{i=1}^N, \{HL_i\}_{i=1}^N, \{HH_i\}_{i=1}^N$, where $LL_N$ contains the low-frequency information at the  $N$-th level,  $LH_i$, $HL_i$ and $HH_i$ contain the horizontal, vertical and diagonal details at the $i$-th level,  respectively.

\begin{figure*}  
	\centering\includegraphics[width=6.3in]{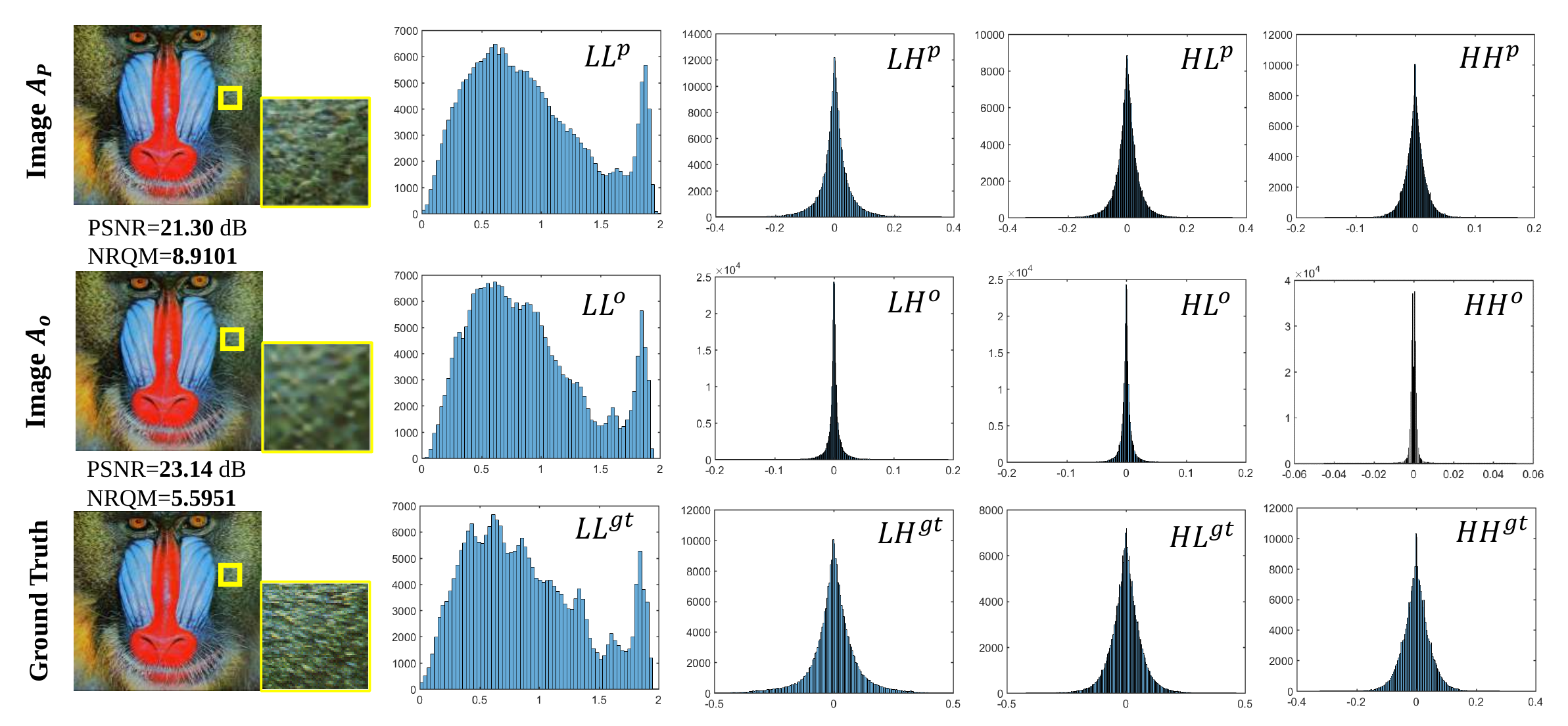}	
	\caption{The first row shows the histograms of different sub-bands of $A_p$ which has high perceptual quality but low objective quality. The second row shows the histograms of different sub-bands of $A_o$ which has high objective quality but low perceptual quality. The third row shows the ground-truth histograms.}
	\label{wave_hist}
	\vspace{-.5em}
\end{figure*}
\begin{figure*}
	\begin{center}
		\subfigure[]{\includegraphics[width=.515\linewidth]{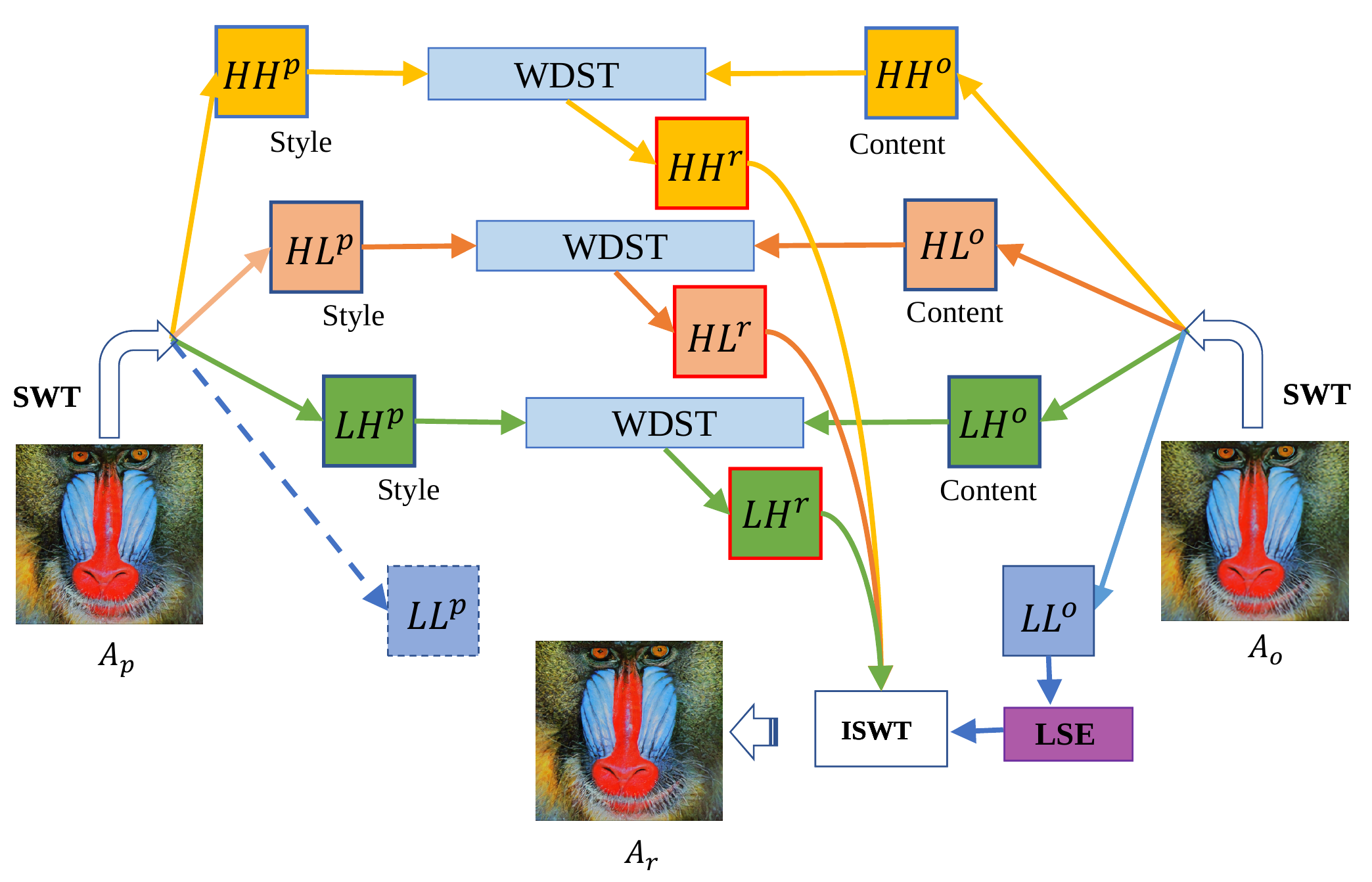}}
		\subfigure[]{\includegraphics[width=.28\linewidth]{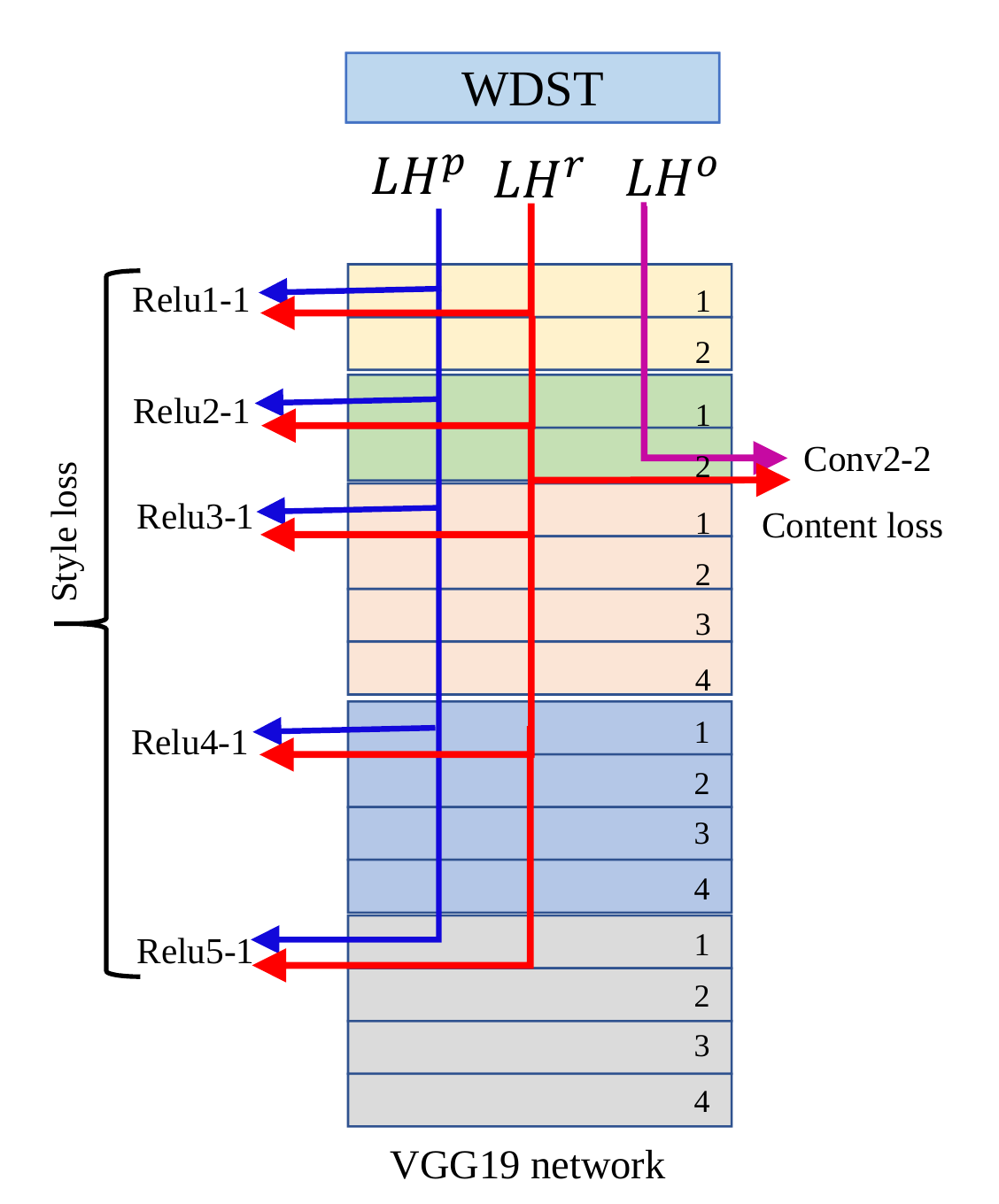}}
		\subfigure[]{\includegraphics[width=.157\linewidth]{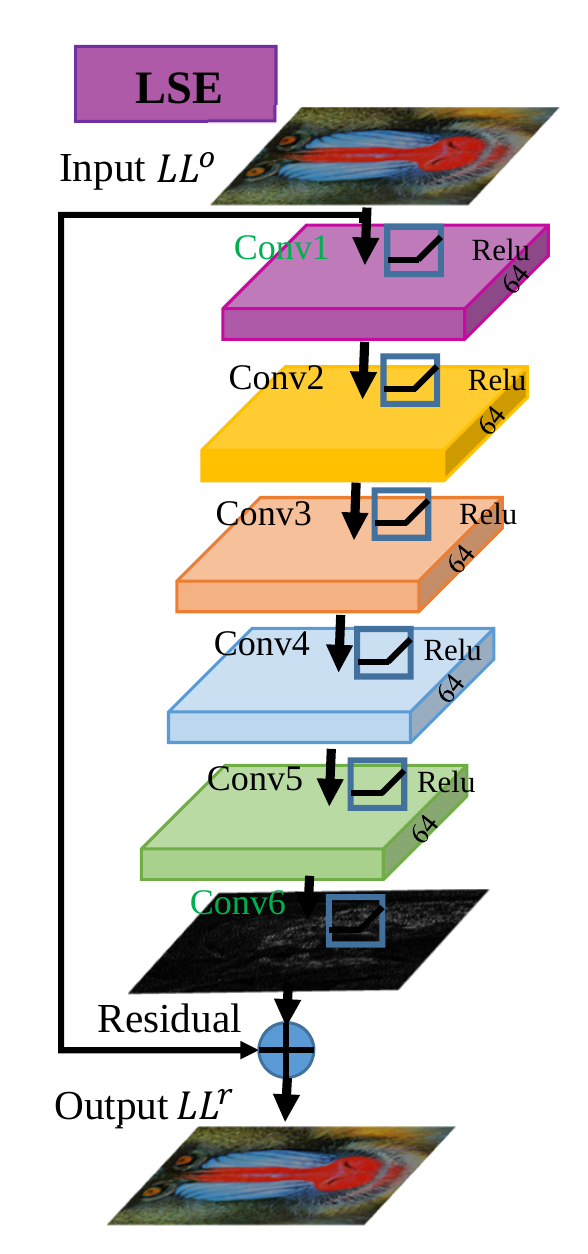}}
	\end{center}
	\vspace{-1em}	
	\caption{(a) shows the framework of our method,  (b) illustrates the wavelet domain style transfer (WDST) algorithm, and (c) shows the low-frequency sub-band enhancement (LSE) network. }\label{fra}
	\vspace{-1em}	
\end{figure*}

\textbf{Motivation. } The 2D SWT can decompose an image into multiple sub-bands, including one low-frequency  and several high-frequency sub-bands. Our key insight here is that the low-frequency sub-band has a significant effect on the objective quality of the image, while the high-frequency sub-bands affect the perceptual quality significantly. To verify that, we consider two super-resolved images: $A_p$ with high perceptual quality but low objective quality,  and $A_o$ with high objective quality but low perceptual quality. Fig. \ref{wave_hist} shows these two images, together with the histograms of their sub-bands after SWT. Here,  $A_p$ and $A_o$ are obtained using the existing SISR methods CX \cite{mechrez2018learning} and EDSR \cite{lim2017enhanced}, respectively.  We use peak signal-to-noise ratio (PSNR)  to measure the objective quality, and  NRQM  \cite{ma2017learning} to measure the perceptual quality following \cite{mechrez2018learning}. Note that larger  PSNR and NRQM values indicate better objective and perceptual quality, respectively. As shown in Fig. \ref{wave_hist},  the high-frequency sub-bands (i.e., LH, HL, HH) of $A_p$ have quite similar histogram distributions as  the ground-truth, but that is not the case for $A_o$. Since the high-frequency sub-bands contain the detail information, this can explain why $A_p$ has high perceptual quality.  For the LL sub-band,  $A_o$ has a more similar histogram  as  the ground-truth than  $A_p$, which is one of the reasons why $A_o$ has high objective quality.

In order to further verify our observation,  a simple substitution experiment is performed as follows. We replace the low-frequency sub-band of $A_p$ with that of $A_o$, and keep all its high-frequency sub-bands. These sub-bands are combined via 2D inverse SWT (ISWT) to obtain a reconstructed image $\tilde{A}_{p}$. Likewise, we replace the low-frequency sub-band of $A_o$ with that of $A_p$ to obtain a reconstructed image $\tilde{A}_{o}$. Table \ref{observe} shows the PSNR and  NRQM results on the BSD100 dataset. As can be seen, the PSNR of $\tilde{A}_{p}$ improves more than 1dB over $A_{p}$ while the NRQM score does not change too much. Similar phenomenon can be observed between $\tilde{A}_{o}$  and $A_{o}$. The reason why the objective quality is significantly affected is that the low-frequency sub-band is changed. In contrast, the perceptual quality is not particularly influenced because we preserve the high-frequency sub-bands. Thus, in order to obtain an image with a good PD tradeoff, one possible solution is to pursue high objective quality of its low-frequency sub-band  and high perceptual quality of its high-frequency sub-bands.

\begin{table}
	\begin{center}
		\caption{PSNR and NRQM scores  on the BSD100 dataset.}\label{observe}
		\begin{tabular}{c|cccc}
			\hline \hline
			Methods&$A_p$&$\tilde{A}_{p}$&$A_o$&$\tilde{A}_{o}$ \\ \hline
			PSNR  &24.58&25.68&27.80& 26.57 \\
			NRQM&8.8007&8.7775&5.7159&5.8864 \\			
			\hline
		\end{tabular}
	\end{center}
	\vspace{-3em}		
\end{table}

\begin{table*}\addtolength{\tabcolsep}{-3.3pt}	
	\begin{center}
		\caption{Benchmark comparisons for 4 $\times$ upscaling,  with the best results bold and the second bests underlined.}\label{result} 		
		\begin{tabular}{c|ccc|cccccc}
			\hline \hline
			Set5            &Bicubic &EDSR\cite{lim2017enhanced} &CX\cite{mechrez2018learning}  &SRGAN-MSE\cite{ledig2017photo} &G–MGBP\cite{michelini2018multi}    &PESR\cite{vu2018perception}   & Deng\cite{deng2018enhancing}&ESRGAN\cite{wang2018esrgan}   &Ours \\ \hline
			PSNR       &28.42   &32.63 &29.10    &30.66     &30.87  &30.76  &\underline{31.14}  &31.11     &\textbf{31.46} \\
			SSIM       &0.8245  &0.9117&0.8523   &0.8758    &0.8807 &0.8915 &\underline{0.8917} &0.8839    &\textbf{0.8929} \\
			NRQM       &3.7624  &5.2106&7.9566   &7.3082    &\underline{7.3115} &7.1344 &7.0022 &7.0724    &\textbf{7.5180} \\			
			\hline
			Set14           &Bicubic &EDSR\cite{lim2017enhanced}  &CX\cite{mechrez2018learning}      &SRGAN-MSE\cite{ledig2017photo} &G–MGBP\cite{michelini2018multi}    &PESR\cite{vu2018perception}    & Deng\cite{deng2018enhancing}  &ESRGAN\cite{wang2018esrgan}  &Ours \\ \hline
			PSNR        &26.10   &28.95 &26.01    &27.01     &27.56  &27.57  &\underline{27.77}  &27.53     &\textbf{28.07} \\
			SSIM        &0.7850  &0.8583&0.7839   &0.8033    &0.8206 &0.8322 &\underline{0.8325} &0.8228    &\textbf{0.8356} \\
			NRQM        &3.6598  &5.3788&7.9423   &\textbf{7.8770}    &7.5042 &7.5301&7.5575 &7.5936    &\underline{7.6827} \\
			\hline
			BSD100          &Bicubic &EDSR\cite{lim2017enhanced}  &CX\cite{mechrez2018learning}      &SRGAN-MSE\cite{ledig2017photo} &G–MGBP\cite{michelini2018multi}    &PESR\cite{vu2018perception}    & Deng\cite{deng2018enhancing}  &ESRGAN\cite{wang2018esrgan}   &Ours \\ \hline
			PSNR      &25.96   &27.80 &24.58    &25.98     &\underline{26.59}  &26.33  &26.46  &26.44     &\textbf{26.82} \\
			SSIM      &0.6675  &0.7432&0.6432   &0.6429    &0.6926 &0.6980 &\underline{0.7048} &0.7002    &\textbf{0.7058} \\
			NRQM      &3.7207  &5.7159&8.8007   &8.4276    &8.1790 &8.3298 &\underline{8.4452} &8.3034    &\textbf{8.5948} \\
			\hline		
			Urban100          &Bicubic &EDSR\cite{lim2017enhanced}  &CX\cite{mechrez2018learning}      &SRGAN-MSE\cite{ledig2017photo} &G–MGBP\cite{michelini2018multi}    &PESR\cite{vu2018perception}    & Deng\cite{deng2018enhancing}  &ESRGAN\cite{wang2018esrgan}   &Ours \\ \hline
			PSNR &23.14   &26.86 &24.00    &- &25.15  &25.88  &25.96 &\underline{26.08} &\textbf{26.26} \\
			SSIM &0.9011  &0.9679&0.9313   &- &0.9495 &0.9610 & 0.9620&\underline{0.9624} &\textbf{0.9649} \\
			NRQM &3.4412  &5.3365&6.7982   &- &6.2190 & 6.3190      & \underline{6.4317}&6.1762 &\textbf{6.4556} \\
			\hline		
			PIRM         &Bicubic &EDSR\cite{lim2017enhanced}  &CX\cite{mechrez2018learning}      &SRGAN-MSE\cite{ledig2017photo} &G–MGBP\cite{michelini2018multi}    &PESR\cite{vu2018perception}    & Deng\cite{deng2018enhancing}  &ESRGAN\cite{wang2018esrgan}   &Ours \\ \hline
			PSNR &26.51   &28.72 &25.41    &- &27.17 &27.11  &\underline{27.48}  &26.66 &\textbf{27.63} \\
			SSIM &0.8232  &0.8930&0.8177   &- &0.8524 &0.8649 &\underline{0.8728} &0.8529 & \textbf{0.8755}\\
			NRQM &3.8376  &5.7116&8.5746   &- &8.0556 &8.2172 &8.1665 &\underline{8.2445} & \textbf{8.3692}\\
			\hline
		\end{tabular}
	\end{center}
	\vspace{-1.5em}
\end{table*}

Fig. \ref{fra} (a) shows the framework of our method. Given one image $A_p$ with high perceptual quality and another image $A_o$ with high objective quality, we first perform 2D SWT on these two images, so that each image is decomposed into one low-frequency and several high-frequency sub-bands. Take the decomposition with one level for example, $A_p$ is decomposed into $\{LL^p,LH^p, HL^p, HH^p\}$, and $A_o$ is decomposed into $\{LL^o,LH^o, HL^o, HH^o\}$.  For $LL_o$, we use LSE network to enhance its objective quality. For high-frequency sub-bands pairs, e.g., $LH_p$ and $LH_o$, we use WDST to fuse them to a new sub-band. Finally, all fused sub-bands and enhanced $LL_o$ are synthesised by ISWT to obtain image $A_r$.

\textbf{Low-frequency sub-band enhancement (LSE). } For the low-frequency sub-band $LL^o$, we aim to further improve its objective quality. Here, we employ the basic network structure of VDSR \cite{kim2016accurate} to achieve this goal, as shown in Fig. \ref{fra} (c). The network is composed of 6 convolutional layers with a rectified linear unit (Relu) after each layer. For each layer, the filter size  is $3\times3$ and the number of filters is 64. The input to the network is the low-frequency sub-band $LL^o$ from the image $A_o$, and the target is the $LL^{gt}$ from the ground-truth image $A_{gt}$. To speed up the training process, we also use the residual learning strategy which learns the difference between target $LL^{gt}$ and the input $LL^o$.  The training goal is to minimize the $\ell_2$ norm between the predicted outputs $LL^r$ and the ground truth $LL^{gt}$:
\begin{equation}
\mathcal{L}=\sum_{i=1}^N \|LL^{gt}(i)- LL^r(i)\|_2,
\end{equation}
where $LL^r$ is the sum of  $LL^o$ and the learned residual map. 

\textbf{Wavelet domain style transfer (WDST). }   For the high-frequency sub-bands, we propose a wavelet domain style transfer (WDST) algorithm to improve their perceptual quality. Take the sub-band pair $LH^p$ and $LH^o$ for example, as shown in Fig. \ref{wave_hist}, the wavelet coefficients in $LH^p$ are richer than those in $LH^o$, i.e., $LH^p$ contains more non-zero wavelet coefficients  than $LH^o$. We wish to transfer the detailed wavelet coefficients in  $LH^p$ to  $LH^o$, so that $LH^o$ can have  higher perceptual quality. Thus, we regard $LH^p$ as the style input and  $LH^o$ as the content input to generate  an output sub-band  $LH^r$ using WDST. Different from the conventional style transfer algorithm where the inputs are pixel values, we use the wavelet coefficients as inputs in the WDST. Since the wavelet coefficients can be negative or larger than 1, a  pre-processing step is required to  normalize them between 0 and 1. 

\begin{figure*}  
	\centering\includegraphics[width=6.8in]{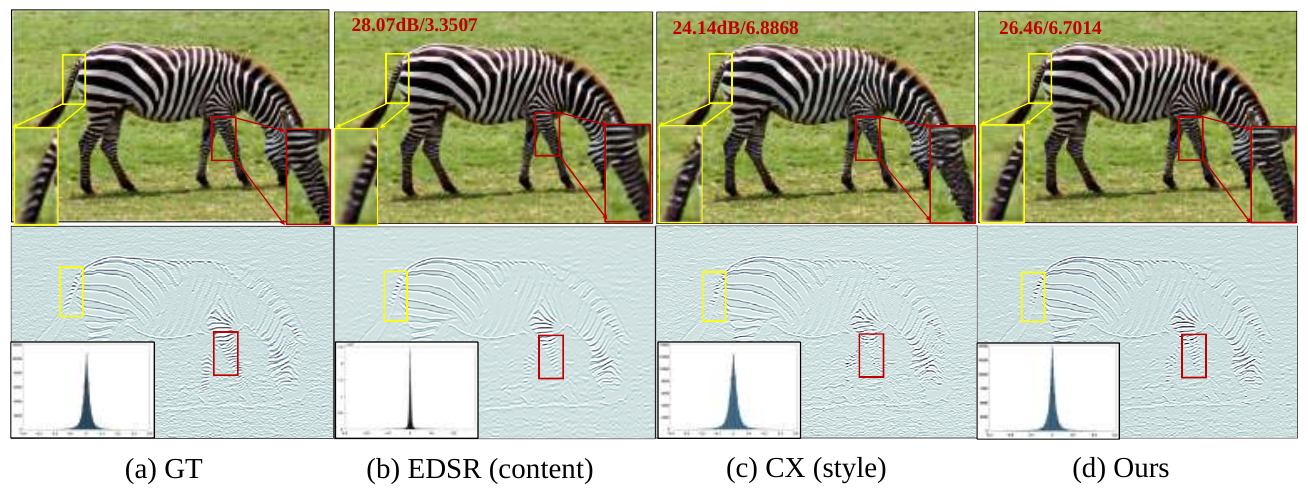}	
	\caption{The first row shows the restored images of \textit{Zebra}  in Set 14 using EDSR, CX and our method, with the red values indicating the PSNR/NRQM values. The second row visualizes the HL sub-bands of the images in the first row, together with  the histograms.  }
	\label{wdst_effect}
	\vspace{-1.5em}
\end{figure*}

After normalization, for each high-frequency sub-band pair, the WDST algorithm is performed  by minimizing a  loss function that combines the content loss $L_c$, style loss $L_s$ \cite{gatys2016image} and a $\ell_1$ norm loss.  The  $\ell_1$ norm loss is specifically added to preserve the sparsity of wavelet coefficients. The total loss function for the $LH$ sub-band is defined as:
\begin{equation} \label{eeww}
\small
\mathcal{L}_{LH}= \alpha L_{c}(LH^r, LH^o)+ \beta L_{s}(LH^r, LH^p)+\hspace{-.2em}\gamma\|LH^r\|_1, 
\end{equation}
where  $\alpha$, $\beta$ and $\gamma$ are the weights for the content, style and  $\ell_1$ norm loss, respectively.  The content loss is defined as the MSE between the feature maps of the content input and the generated output at a specific layer $L$ of a pre-trained VGG network \cite{simonyan2014very}:
\begin{equation}
\small
L_{c}(LH^r, LH^o)=\frac{1}{2\sqrt{N_LM_L}}\sum_{i,j}(F_{ij}^L(LH^r)-F_{ij}^L(LH^o))^2.
\end{equation}
Here, $F^L(LH^r)$ and $F^L(LH^o)$ are the feature maps at layer $L$ of a pre-trained VGG network \cite{simonyan2014very} with $LH^r$ and $LH^o$ as inputs, respectively. In addition, $N_L$ is the number of feature maps  at layer $L$, and $M_L$ is the product of the width and height of the feature map. Different from the content loss which is calculated between $LH^o$ and $LH^r$, the style loss is calculated between the style input $LH^p$ and $LH^r$. Moreover, unlike the content loss calculated at a single layer,   the total style loss is defined by a weighed sum of the style loss at different layers:
\begin{equation}
\small
L_{s}(LH^r, LH^p)=\sum_l w_lL_{s}^l(LH^r, LH^p),
\end{equation}
where $w_l$ is the weight for the style loss at the $l$-th layer. The $L_{s}^l(LH^r, LH^p)$ is calculated as the MSE between the Gram matrices of feature maps at the $l$-th layer in the pre-trained VGG network with $LH^r$ and $LH^p$ as inputs, respectively. Mathematically, it is defined as: 
\begin{equation}
\small
L_{s}^l(LH^r, LH^p)=\frac{1}{4N_l^2M_l^2}\sum_{ij}(G_{ij}^l(LH^r)-G_{ij}^l(LH^p))^2,
\end{equation}
where the $G^l(LH^r)$ and $G^l(LH^p)$ are the Gram matrices at the $l$-th layer for $LH^r$ and $LH^p$, respectively. We have $G^l(LH^r)=F^l(LH^r)^TF^l(LH^r)$ and $G^l(LH^p)=F^l(LH^p)^TF^l(LH^p)$.  The layer conv 2-2 in VGG network \cite{simonyan2014very} is used to calculate the content loss, and layers Relu1-1, Relu2-1, Relu3-1, Relu4-1, and Relu5-1 are used to calculate the style loss. With all loss defined, following \cite{gatys2016image}, we use L-BFGS algorithm \cite{zhu1997algorithm} to obtain  $LH^r$ in \eqref{eeww} in a gradient decent way.  Similarly,  we can obtain  $HL^r$ and  $HH^r$.

After obtaining high-frequency sub-bands  $LH^r$, $HL^r$, and  $HH^r$,  we need to de-normalize them. Then,  
we can reconstruct  image $A_r$  by performing 2D ISWT on these high-frequency sub-bands together with the low-frequency sub-band $LL^r$ using the synthesis low-pass and high-pass filters $H_1$ and $G_1$. Here, for perfect reconstruction, $H_1$ and $G_1$ are the synthesis wavelet filters related to the analysis filters $H_0$ and $G_0$ used in the decomposition \cite{mallat1989theory}.

\begin{figure}  
	\centering\includegraphics[width=3.3in]{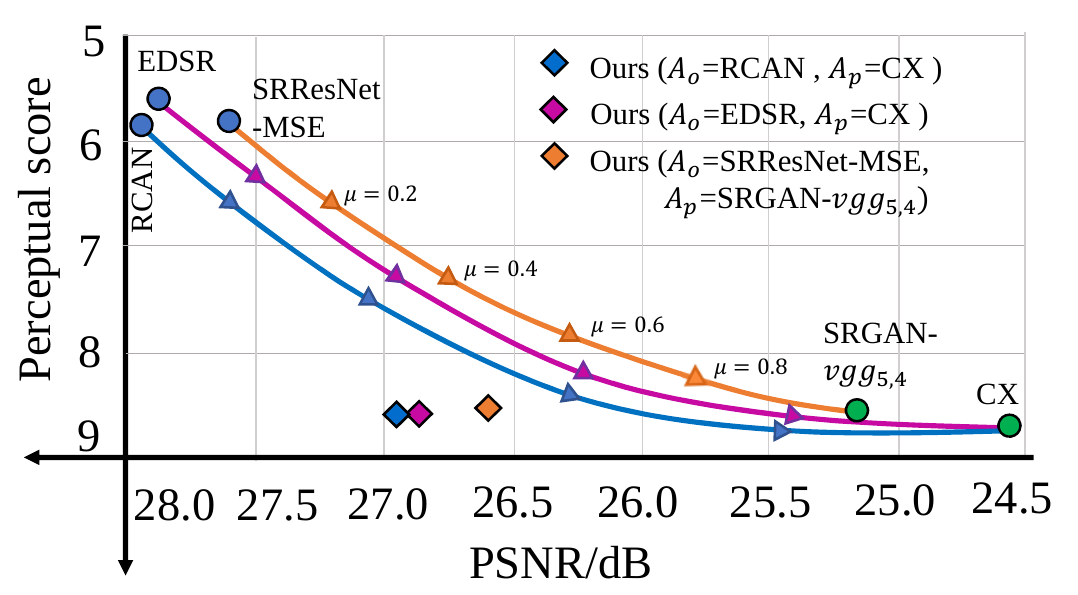}
	\caption{The perception-distortion (PD) curves of EDSR and CX, RCAN and CX, SRResNet-MSE and SRGAN-vgg54.}
	\vspace{-1.5em}
	\label{curve}	
\end{figure}
\begin{figure*}  
	\centering\includegraphics[width=6.8in]{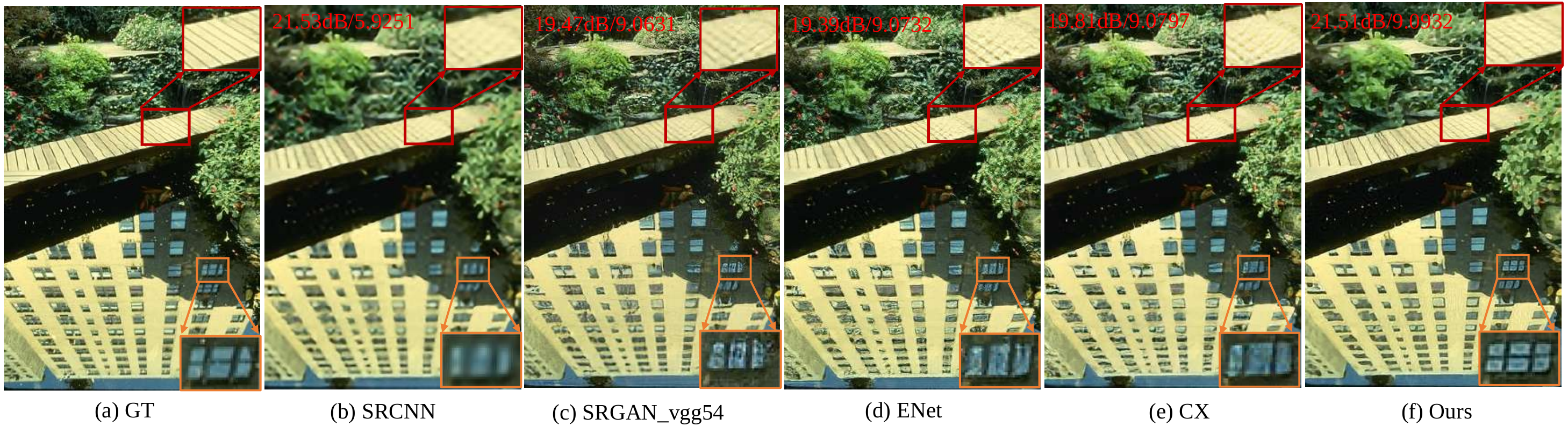}	
	\caption{Visual comparisons of  image  from BSD100 for 4$\times$ upscaling. The red numbers indicate the PSNR and NRQM values.  }
	\label{building}
	\vspace{-.7em}
\end{figure*}

\begin{figure*}  
	\centering\includegraphics[width=6.8in]{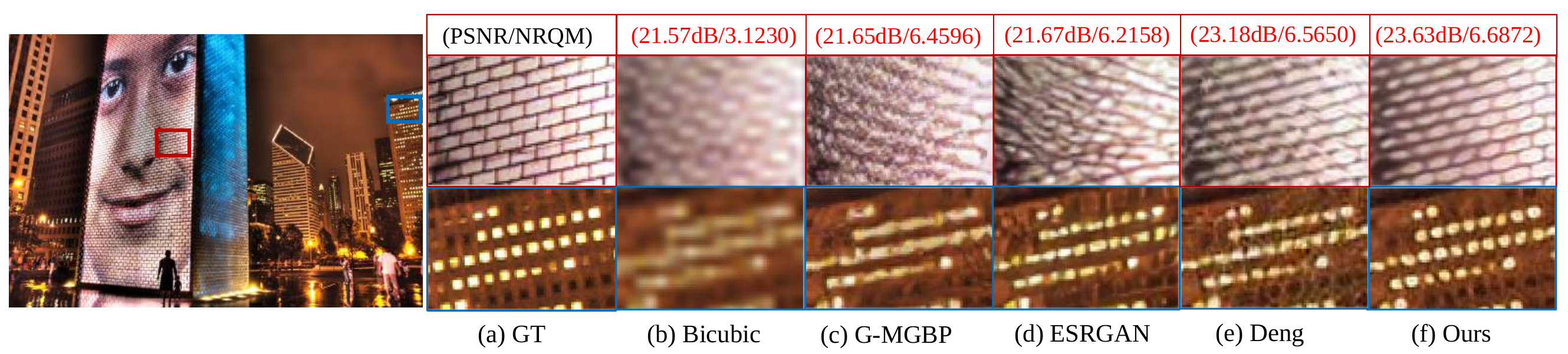}	
	\caption{Visual comparisons of  image  from Urban100 for 4$\times$ upscaling. The red numbers indicate the PSNR and NRQM values.  }
	\label{vase}
	\vspace{-.5em}
\end{figure*}
\begin{figure*}
	\begin{center}
		\subfigure[SRGAN-MSE \cite{ledig2017photo} and ours]{\includegraphics[width=.49\linewidth]{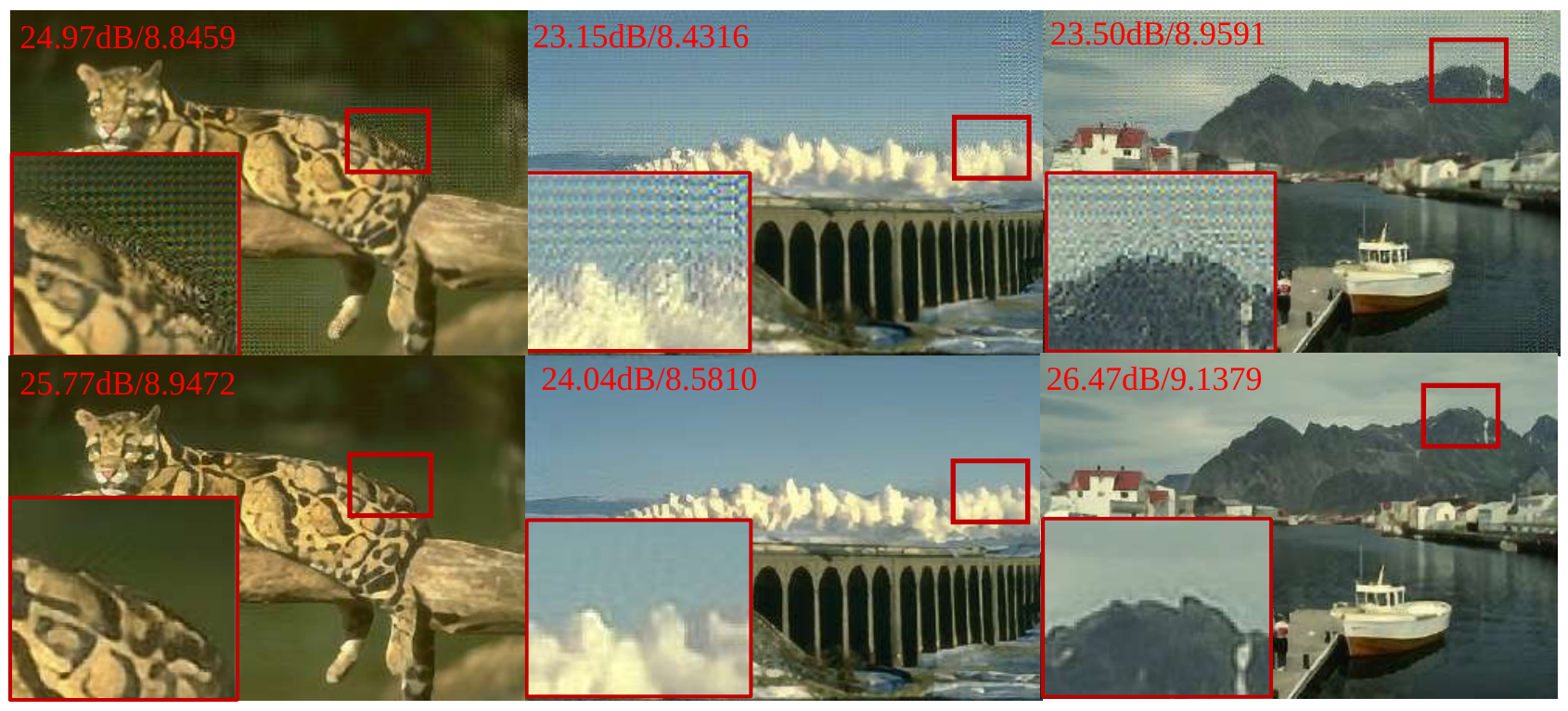}}
		\subfigure[Deng \cite{deng2018enhancing} and ours]{\includegraphics[width=.49\linewidth]{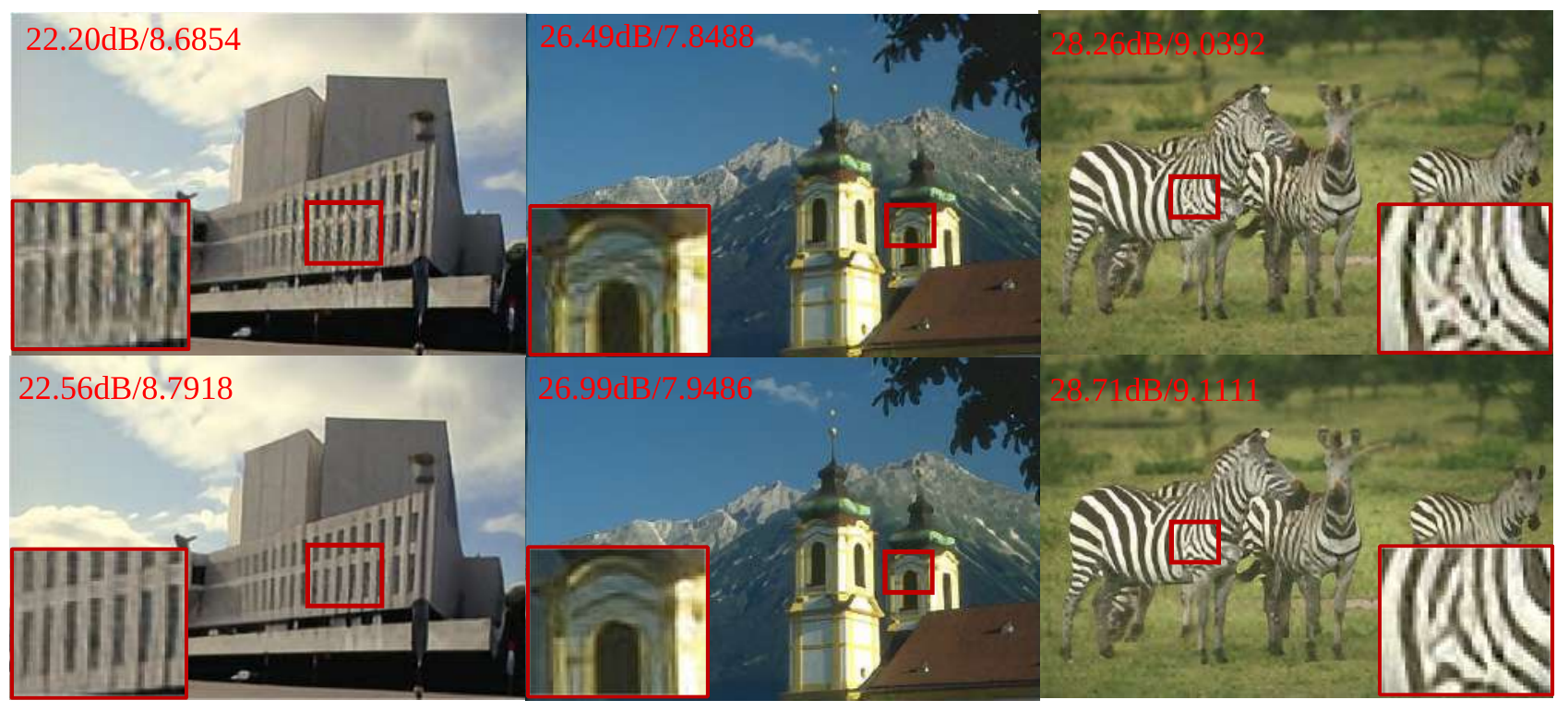}}
	\end{center}
	\vspace{-.5em}	
	\caption{(a) compares the images between SRGAN-MSE and ours, (b) compares the images of Deng and ours. The first rows in (a) and (b) are SRGAN-MSE and Deng, and the second row is our method. The red numbers indicate the PSNR and NRQM values. }\label{frra}	
	\vspace{-1.2em}
\end{figure*}

\section{Numerical results} \label{er}

\textbf{Experimental setup.} For the 2D SWT, we use $bior2.2$ as the default wavelet filter. The number of wavelet decomposition levels is  2, which means we have six high-frequency sub-bands and one low-frequency sub-bands (see Fig. \ref{swt}). In the LSE process, the loss function is minimized using the  stochastic gradient descent (SGD) with backpropagation. The batch size is 64, the basic learning rate is 0.01 and the momentum is 0.9.  In the WDST process, the ratio between the content loss and the style loss is $10^{-3}$,  the ratio between the content loss and the $\ell_1$ norm  loss is $10^{-5}$, and the weight of each layer when calculating the style loss is  0.2. The maximum iteration number is 5000 and 1000 for the first and second level decompositions, respectively.  We use EDSR method \cite{lim2017enhanced} to obtain $A_o$, and CX method \cite{mechrez2018learning} to obtain $A_p$.   Following \cite{mechrez2018learning}, the perceptual score is calculated using  NRQM  \cite{ma2017learning}. We evaluate the performance of our method on various datasets, including Set5 \cite{bevilacqua2012low}, Set14 \cite{zeyde2010single}, BSD100 \cite{martin2001database}, Urban100 \cite{huang2015single}, and PIRM \cite{blau20182018}.

\textbf{Benchmarks.}  The comparison methods are classified into three categories: methods that aim to improve the objective quality including A+ \cite{timofte2014a+}, Self-Ex \cite{huang2015single}, SRCNN \cite{dong2014learning}, ESPCN \cite{shi2016real}, SRResNet-MSE \cite{ledig2017photo}, VDSR \cite{kim2016accurate}, EDSR \cite{lim2017enhanced}, and RCAN \cite{zhang2018image}; methods that aim to improve the perceptual quality including SRGAN-vgg54 \cite{ledig2017photo}, SRGAN-vgg22\cite{ledig2017photo}, ENet \cite{sajjadi2017}, and CX \cite{mechrez2018learning}; and methods that aim to improve both the objective and perceptual quality including SRGAN-MSE \cite{ledig2017photo}, G–MGBP\cite{michelini2018multi}, PESR \cite{vu2018perception}, EUSR\cite{choi2018deep},  Deng \cite{deng2018enhancing}  and ESRGAN\cite{wang2018esrgan}.

\textbf{Effectiveness of WDST.} In order to show the effectiveness of our WDST algorithm, we visualize in Fig. \ref{wdst_effect} the input content and style sub-bands, as well as the output sub-band using the WDST algorithm. As can be seen, the content sub-band lacks many high-frequency details and the style sub-band has  messy structures, e.g, the horse leg and tail.  After the WDST, the output sub-band overcomes  these drawbacks, which  is now abundant in high-frequency details and has clear textures and structures.  To some extent, the output sub-band corrects the wrong information in the style sub-band and re-locate it in the right place, with the guidance of the content sub-band. We also show the histogram distributions of the sub-bands in Fig. \ref{wdst_effect}. It can be seen that our histogram is closer to the ground-truth compared to EDSR, which is the reason why we have higher perceptual quality.

\textbf{Wavelet filter sensitivity.} In our algorithm, we use wavelet filter to decompose each image into various sub-bands.
\begin{table}\addtolength{\tabcolsep}{-4pt}	
	\small
	\begin{center}
		\caption{Effects of wavelet filter on Set 14 dataset.}\label{ore}
		\vspace{-.5em}	
		\begin{tabular}{c|ccccccc}
			\hline \hline
			Filter& $haar$& $db2$&$bior2.2$ & $rbior2.2$&$coif2$&$db4$&$bior4.4$ \\ \hline			
			PSNR&28.06 &\textbf{28.08} &28.07 &27.96 &28.05 &28.06 &28.05  \\
			SSIM&\textbf{0.8379} &0.8369 &0.8356 &0.8336 &0.8344 &0.8348 &0.8343\\
			NRQM&7.5109 &7.6103 &7.6827 &7.6403 &7.7101 &7.6928 &\textbf{7.7442} \\		\hline			
			\hline
		\end{tabular}
	\end{center}
	\vspace{-2.5em}		
\end{table}
\begin{table}\addtolength{\tabcolsep}{-1.5pt}	
	\begin{center}
		\caption{Ablation study of WDST on each sub-band.}\label{ab}	
		\begin{tabular}{c|ccc|ccc}
			\hline \hline
			Sub-band	&$LH$ &$HL$ &$HH$ &PSNR& SSIM& NRQM \\ \hline
			\multirow{4}*{WDST}&N&Y&Y&27.19&0.7195&7.8490  \\
			&Y&N&Y&27.28&0.7227&7.8343 \\
			&Y&Y&N&26.96&0.7105&8.0542\\	
			&Y&Y&Y&26.82&0.7058&8.5948\\		
			\hline
		\end{tabular}
	\end{center}
	\vspace{-2.5em}		
\end{table}
In order to investigate the effects of wavelet filter on the performance of our algorithm, we present in Table \ref{ore} the PSNR, SSIM and NRQM results with different wavelet filters. These filters include $haar$, $db2$ and $db4$ from Daubechies, $bior2.2$ and $bior4.4$ from Biorthogonal,  $rbio2.2$ from Reverse biorthogonal, and $coif2$ from Coifman wavelet family. From Table \ref{ore}, we can see that the wavelet filter indeed has some effects on the performance. Specifically, the $haar$ filter has the highest SSIM value, the $db2$ filter  performs best in PSNR and the $bior4.4$ filter has the best perceptual quality. However, the difference among different filters is not very significant.  

\textbf{Perception-distortion (PD) performance.} Fig. \ref{comp_build} compares the PD performance of different methods in  the PSNR and NRQM plane. As we can see, methods A+, Self-Ex, SRCNN, ESPCN, SRResNet-MSE, VDSR, EDSR, RCAN occupy the upper left region which means they have high objective quality but low perceptual quality. In contrast,  methods SRGAN-vgg54, SRGAN-vgg22, ENet, and CX take up the bottom right region, which indicates they have high perceptual quality but low objective quality. Other methods like SRGAN-MSE, PESR, Deng, and ESRGAN stand in the middle region, which are all trying to achieve a good tradeoff between  distortion and perceptual quality. Among all these methods, our method is the closest to the bottom left corner, which means that we achieve the best trade-off between the objective and perceptual quality.
Table \ref{result} compares the numerical results of our method with SRGAN-MSE \cite{ledig2017photo}, G–MGBP\cite{michelini2018multi} , PESR \cite{vu2018perception},  Deng \cite{deng2018enhancing}  and ESRGAN\cite{wang2018esrgan} (with $\alpha =$0.8), which all aim to improve both the perceptual and objective quality. As we can see, our method  outperforms others in both  perceptual and objective quality.

\textbf{Content and Style inputs sensitivity.}
To show the position of our method more clearly, we draw in Fig. \ref{curve} the PD curve of EDSR and CX, which are the two default methods  to generate $A_o$ and $A_p$ in this paper. The curve is drawn by interpolating the pixel values of $A_o$ and $A_p$ with a parameter $\mu \in [0,1]$, as follows 
\begin{equation}
A_r=\mu*A_p+(1-\mu)*A_o.
\vspace{-0.5em}
\end{equation}
Obviously, when  $\mu$ increases, the NRQM  increases while the PSNR decreases. As we can see from Fig. \ref{curve}, our method is far lower than that PD curve, which means we are much better than the simple interpolation of $A_o$ and $A_p$. To investigate our sensitivity to the content and style inputs, we also draw the PD curves of RCAN \cite{zhang2018image} and CX,  SRResNet-MSE and SRGAN-vgg54 \cite{ledig2017photo}, together with our correspoding results.  We can see that, even in the worst case (with SRResNet-MSE and SRGAN-vgg54 as inputs), our algorithm still achieves better PD trade-off (i.e., PSNR/NRQM=26.56 dB/8.5005) than Deng (26.46 dB/8.4452) and ESRGAN (26.44 dB/8.3034).

\textbf{Visual comparison.} 
Figs. \ref{building} and \ref{vase} visualize the images of  our and other methods.   We can see from Fig. \ref{building} that our method can restore correctly the texture of the bridge and the structure of the window, while others either distort the texture or struggle to restore the structure. From Fig. \ref{vase}, we can see that our method can restore the wall and lights clearly, while others fail to do so. 
Our method also overcomes many drawbacks of other methods. Fig. \ref{frra} (a) compares our method with SRGAN-MSE \cite{ledig2017photo}. We can see that the SRGAN-MSE method produces lots of abnormal noise and wrong textures in the images, while our method does not have these problems. Fig. \ref{frra} (b) compares our method with Deng \cite{deng2018enhancing}, which shows  that the images of  Deng \cite{deng2018enhancing} are noisy and have messy structures. In contrast, our method is able to reconstruct images with clean and accurate structures.

\textbf{Ablation study.} In order to study the effects of each high-frequency sub-band on the perception-distortion performance, we show in Table \ref{ab} the results when WDST is not performed on one of the sub-bands. From this table, we can see that each sub-band contributes to the perception-distortion performance. When WDST is absent from any of them, the perceptual quality (NRQM)  decreases significantly. However, compared with LH and HL sub-bands, the influence of HH sub-band is not very significant. This is because the HH sub-band contains the diagnonal information, which is not as much as the horizontal and vertical information contained in the LH and HL sub-bands, respectively.

\vspace{-0.5em}
\section{Conclusion and future work} \label{con}
\vspace{-0.5em}
In this paper,  we have proposed a novel  method based on wavelet domain style transfer, to give an excellent solution to the perception-distortion conflict in SISR. We find that the objective and perceptual quality are influenced by different elements of an image.  To achieve the best trade-off between them, we use stationary wavelet decomposition (SWT) to split elements related with objective quality from those  related with perceptual quality. Then, we can optimize each with different targets,  with little influence on the other.  This “divide and conquer” strategy was demonstrated to achieve a good trade-off between the image distortion and perception, and we
believe this  can inspire more follow-up works to  further push forward the reconstruction performance in SISR. Like the conventional style transfer work \cite{ledig2017photo}, we need many iterations to solve the optimization problem in \eqref{eeww}, which is a little bit time-consuming, i.e., around 60 seconds for each sub-band. Inspired by the real-time artistic style transfer work \cite{johnson2016perceptual}, our future work is to train a feed-forward network to predict the fused sub-band which minimizes \eqref{eeww}, so that the computational complexity can be significantly decreased.

{\small
	\bibliographystyle{ieee_fullname}
	\bibliography{sig}
}

\end{document}